\documentclass{iau} 

\usepackage{graphicx}
\usepackage{color}
\usepackage[colorlinks=true,citecolor=blue,linkcolor=blue]{hyperref}

\title[The merger-driven evolution of ETGs] 
{The merger-driven evolution of \\ massive early-type galaxies}

\author[Carlo Cannarozzo et al.]   
{Carlo Cannarozzo$^{1,2}$,
 Carlo Nipoti$^1$,
 Alessandro Sonnenfeld$^3$,
 Alexie Leauthaud$^4$,
 Song Huang$^5$,
 Benedikt Diemer$^6$,
 Grecco Oyarzún$^4$}

\affiliation{$^{1}$Dipartimento di Fisica e Astronomia, Alma Mater Studiorum Università di Bologna, \\ Via Piero Gobetti 93/2, I-40129 Bologna, Italy\\ email: {\tt carlo.cannarozzo3@unibo.it} \\[\affilskip]
$^{2}$INAF - Osservatorio di Astrofisica e Scienza dello Spazio di Bologna, \\ Via Piero Gobetti 93/3, I-40129 Bologna, Italy\\
$^{3}$Leiden Observatory, Leiden University, \\ Niels Bohrweg 2, 2333 CA Leiden, The Netherlands\\
$^{4}$ Department of Astronomy and Astrophysics, University of California, Santa Cruz, \\ 1156 High Street, Santa Cruz, CA 95064 USA \\
$^{5}$ Department of Astrophysical Sciences, Princeton University, \\ 4 Ivy Lane, Princeton, NJ 08544, USA \\
$^{6}$ NHFP Einstein Fellow, Department of Astronomy, University of Maryland, \\ College Park, MD 20742, USA}
\pubyear{2020}
\volume{359}  
\jname{Galaxy evolution and feedback across different environments}
\editors{T. Storchi-Bergmann, R. Overzier, W. Forman \& R. Riffel, eds.}
\begin{document}

\maketitle

\begin{abstract}
The evolution of the structural and kinematic properties of early-type galaxies (ETGs),
their scaling relations, as well as their stellar metallicity and age contain precious information on the assembly history of these systems.
We present results on the evolution of the stellar mass-velocity dispersion relation of ETGs, focusing in particular on the effects of some selection criteria used to define ETGs. We also try to shed light on the role that in-situ and ex-situ stellar populations have in massive ETGs, providing a possible explanation of the observed metallicity  distributions. 

\end{abstract}

\firstsection 
\section{Introduction}
In the currently favoured model of galaxy formation, early-type galaxies (ETGs) are believed to assemble in two phases (\cite{Oser2010}).
The first phase  ($z\gtrsim2$) is dominated by the \textit{in-situ} star formation. Afterwards, as a consequence of mostly dissipationless minor and major mergers, ETGs accrete stars formed \textit{ex situ}.
This scenario leads to intriguing questions including how the properties of ETGs evolve across cosmic time, whether and to what extent mergers modify the scaling relations of these massive galaxies and shape the distribution of the stellar populations within them.

In this proceeding we present some aspects of the evolution of massive ETGs based on two different works. In  \autoref{sec:mstar_sigma} we report the results on the $M_*$-$\sigma_{\rm e}$ relation of ETGs obtained in \cite{CSN20} (hereafter CSN), and study the effect of adopting two different selection criteria for ETGs.
In \autoref{sec:insitu_exsitu} we present preliminary results of a forthcoming paper (\cite{Cannarozzo2020}) aimed at studying the radial distributions of in-situ and ex-situ stellar components of massive ETGs.

\section{The evolution of the stellar mass-velocity dispersion relation}
\label{sec:mstar_sigma}



The central stellar velocity dispersion $\sigma_\mathrm{e}$ of present-day ETGs is found to correlate with their stellar mass $M_*$. There are indications that this correlation evolves with redshift in the sense that, at given $M_*$, higher-$z$ ETGs have, on average, higher $\sigma_\mathrm{e}$ (e.g., \cite{Tanaka2019}). However, the detailed evolution of the $M_*$-$\sigma_\mathrm{e}$ relation is hard to determine, because of the difficulty of measuring $\sigma_\mathrm{e}$ in large samples of ETGs at high $z$.

CSN studied the evolution of the $M_*$-$\sigma_\mathrm{e}$ relation in massive ($\log (M_*/\mathrm{M_\odot})>10.5$) ETGs in the redshift range $0<z<2.5$, using a Bayesian hierarchical approach. CSN considered a sample of galaxies composed by two main subsamples. The first subsample, named \textit{fiducial sample}, consists of  ETGs in the redshift range $0\lesssim z\lesssim1$ drawn from  the Sloan Digital Sky Survey (SDSS, \cite{Eisenstein2011}) and the Large Early Galaxy Astrophysics Census (LEGA-C, \cite{vanderWel2016}; \cite{Straatman2018}), homogeneously selected  by performing a one-by-one visual inspection to include only objects with elliptical morphology and by applying a cut in the equivalent width (EW) of the emission line doublet [OII]$\lambda\lambda3726,3729$, $\mathrm{EW}([\mathrm{OII}])\geq-5$Å. The second subsample, named \textit{high-redshift sample}, is a more heterogeneous collection of ETGs in the redshift range $0.8\lesssim z\lesssim 2.5$ from previous works in literature (see CSN and references therein).
CSN found that, for both the fiducial and the \textit{extended} (fiducial + high-redshift) samples, the   $M_*$-$\sigma_\mathrm{e}$ relation is well described by $\sigma_\mathrm{e}\propto M_*^\beta(1+z)^\zeta$ with  intrinsic scatter $\simeq 0.08$ dex in $\sigma_{\rm e}$ at given $M_*$ and either $\beta\simeq0.18$ independent of $z$ or redshift-dependent $\beta$ with  $\mathrm{d}\beta/\mathrm{d}\log(1+z)\simeq 0.26$ for the fiducial sample and $\simeq0.18$ for the extended sample; $\zeta$, which measures the redshift dependence of $\sigma_{\rm e}$ at given $M_*$, is $\simeq0.4$ for the fiducial sample  ($0\lesssim z\lesssim1$) and $\simeq0.5$ for  the extended sample ($0\lesssim z\lesssim2.5$).



One of the properties of ETGs is to be passive and EW([OII]) is only one of the possible diagnostics for the star formation rate.
Another indicator is the position of galaxies within the $UVJ$ colour-colour diagram, in which the loci of passive and star-forming galaxies are separate (e.g., \cite{CFN2019}).
For instance, \cite{Belli2014}, from which part of the galaxies of the high-redshift sample are taken, select using a $UVJ$-based criterion. In principle, this different selection criterion can induce spurious evolution when the extended sample is considered.
Here we analyse the effect of adding a $UVJ$-based selection to the criteria used for the fiducial sample.


\begin{figure}[b]
    \centering
        \includegraphics[width=0.35\linewidth]{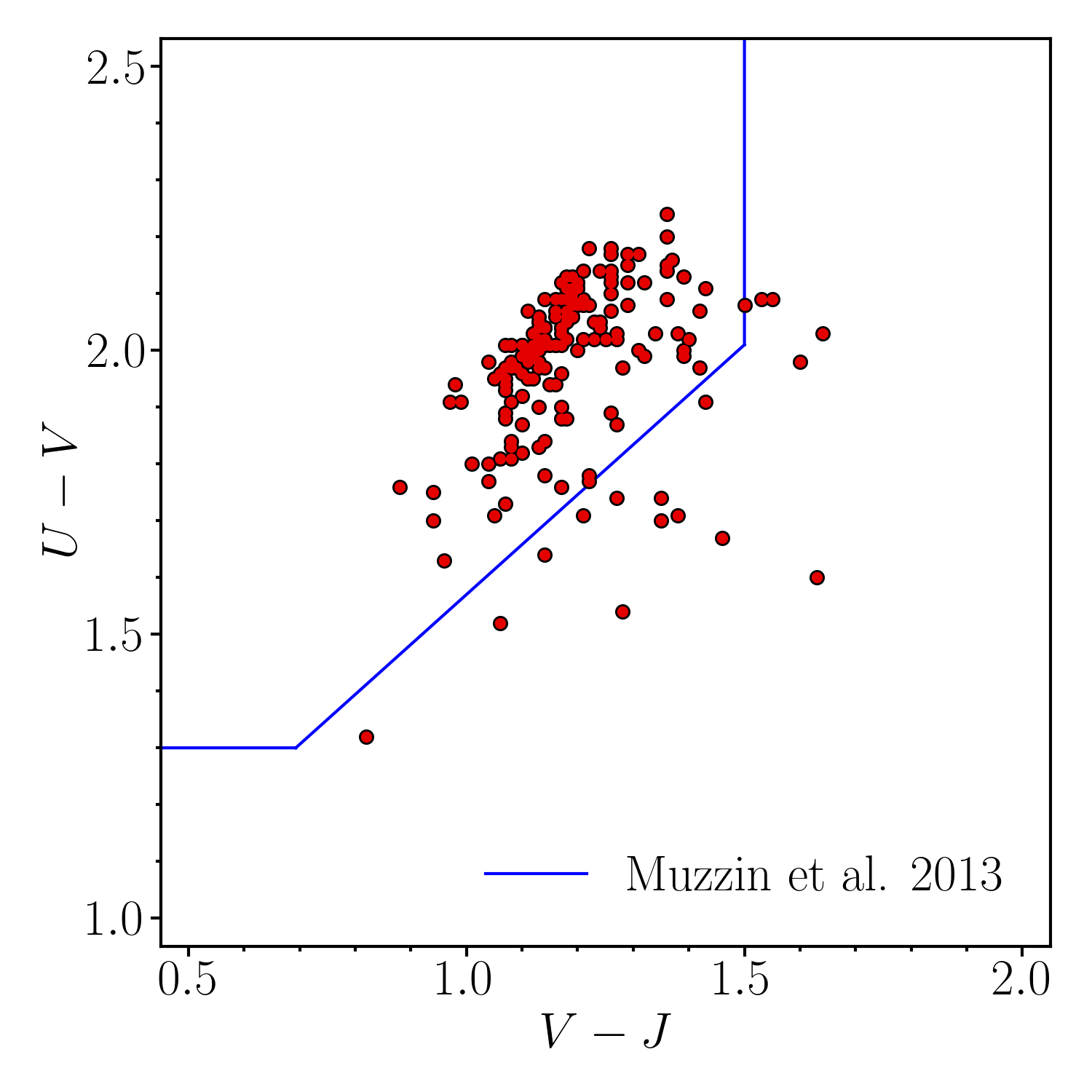}
    \centering
        \includegraphics[width=0.8\linewidth]{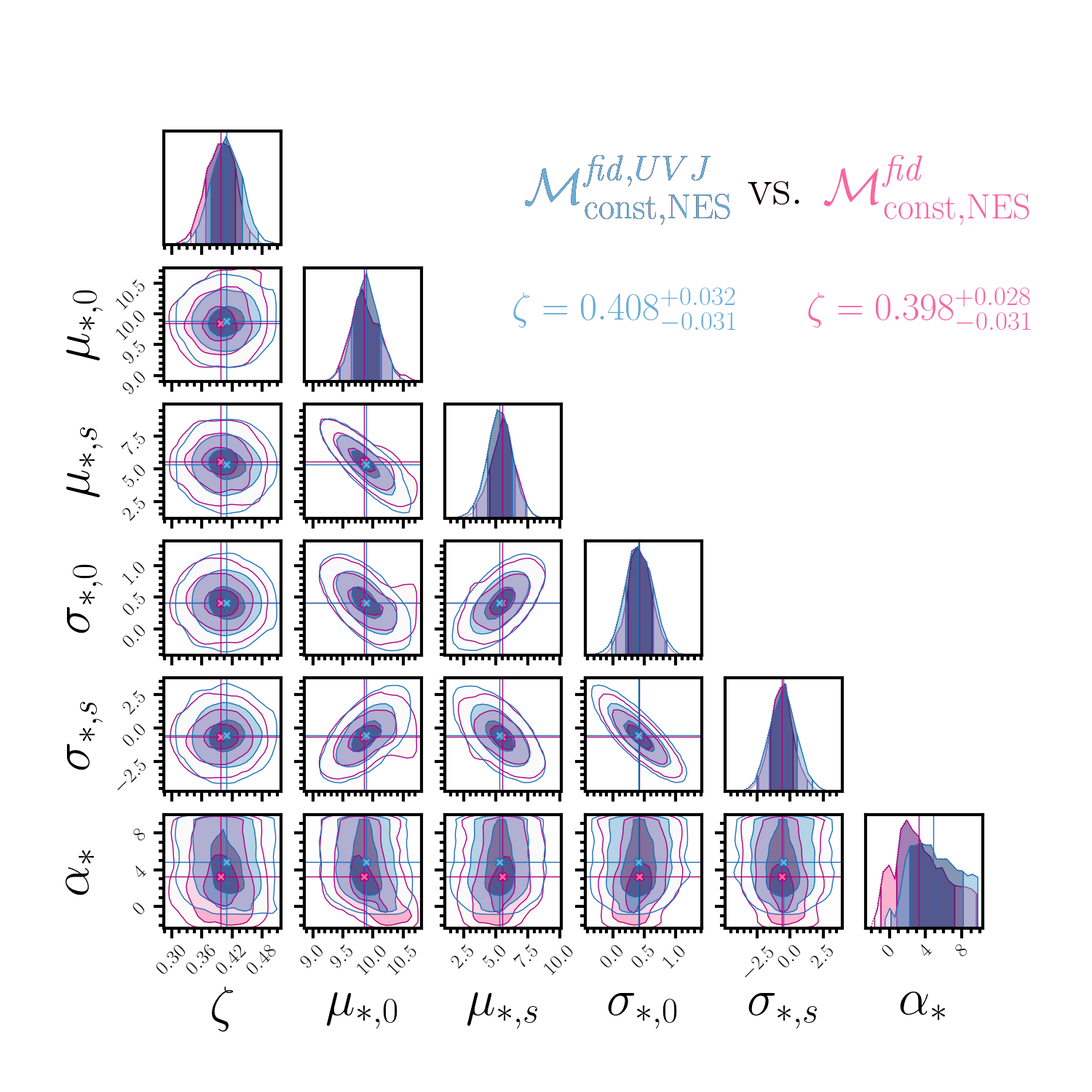}
        \caption{\emph{Top panel:} $UVJ$ colour-colour diagram for the LEGA-C sample of 178 ETGs (red dots). The broken line separates quiescent (upper-left region) and star-forming galaxies (lower-right region) as in \cite{Muzzin2013}.
        \emph{Bottom panel:} 
        posterior probability distributions of the hyper-parameters for the $M_*$-$\sigma_{\rm e}$ models $\mathcal{M}_\mathrm{const,NES}^{fid}$ (purple contours) and $\mathcal{M}_\mathrm{const,NES}^{fid, UVJ}$ (blue contours). In the 1D distributions (top panel of each column) the vertical solid lines and colours delimit the 68, 95 and 99.7-th quantile based posterior credible interval. In the 2D distributions (all the other panels) the contours enclose the 68, 95 and 99.7 percent posterior credible regions. The lines indicate the median values of the hyper-parameters.}
    \label{fig:comparison}
\end{figure}

The model with the highest value of Bayesian evidence explored by CSN, named model $\mathcal{M}_\mathrm{const,NES}$,
has six hyper-parameters: $\zeta$, $\mu_{*,0}$, $\mu_{\rm *,s}$, $\sigma_{*,0}$, $\sigma_{\rm *,s}$ and $\alpha_*$ (see CSN for details). We repeated the analysis of CSN by considering a modified fiducial sample. In particular, we changed the selection criterion for the galaxies of the LEGA-C sample: in addition to the criteria used in CSN, we exclude galaxies that are star-forming on the basis of their position in the $UVJ$ colour-colour diagram.  In the top panel of \autoref{fig:comparison}, the $UVJ$ diagram for the LEGA-C sample of 178 ETGs used in CSN is shown (the $UVJ$ colours are taken from the UltraVISTA catalogue of \textcolor{blue}{\cite[Muzzin \etal\ 2013]{Muzzin2013}}). In this diagram the locus of passive galaxies is the area above and to the left of the broken line:
about $90\%$ of the galaxies of the LEGA-C sample of CSN are in this area. Excluding galaxies that are outside the locus of passive galaxies in the $UVJ$ diagram of \autoref{fig:comparison}, we end up with a modified fiducial sample, consisting of 161 instead of 178 LEGA-C galaxies, in addition to the SDSS galaxies. We applied model $\mathcal{M}_\mathrm{const,NES}$ to this modified fiducial sample (hereafter model $\mathcal{M}_\mathrm{const,NES}^\textit{fid,UVJ}$) and compared the results with those obtained by CSN for the fiducial sample (hereafter model $\mathcal{M}_\mathrm{const,NES}^\textit{fid}$). 
The posterior distributions of the hyper-parameters of  models $\mathcal{M}_\mathrm{const,NES}^\textit{fid,UVJ}$ and  $\mathcal{M}_\mathrm{const,NES}^\textit{fid}$, shown in the bottom panel of \autoref{fig:comparison}, are in agreement within $1\sigma$. In particular, for  model $\mathcal{M}_\mathrm{const,NES}^{fid,UVJ}$, the normalisation of the $M_*$-$\sigma_\mathrm{e}$ scaling relation evolves with $\zeta=0.408_{-0.031}^{+0.032}$, consistent with $0.398_{-0.031}^{+0.028}$ obtained by model $\mathcal{M}_\mathrm{const,NES}^{fid}$. This analysis suggests that, at least as far as the $UVJ$ selection is concerned, the results of the extended sample in CSN are not biased.

\section{In-situ and ex-stu stellar populations in ETGs}\label{sec:insitu_exsitu}
The evolution of metallicity, chemical abundances as well as the ages and other physical properties of stars in ETGs contain information on the evolutionary processes occurred across cosmic time.
In this section we discuss the radial distribution of stellar metallicity in massive MaNGA (\cite{Bundy2015}) ETGs in terms of in-situ and ex-situ stellar components, relying on simulated galaxies drawn from the magnetohydrodynamic cosmological simulation IllustrisTNG (\cite{Springel2018}; \cite{Pillepich2018a}; \cite{Nelson2018}; \cite{Marinacci2018}; \cite{Naiman2018}).


We extracted from the MaNGA survey a sample of more than 700 ETGs, with $M_*\geq10^{10.5}\,\mathrm{M_\odot}$, selected in $\mathrm{EW}(\mathrm{H}\alpha)>-3$Å and Sérsic index $n>2.5$. In order to reduce the effects of systematic biases caused by different assumptions, priors and fitting methods (\cite{Conroy2013}), we rely on two estimates of metallicity derived by using the spectral fitting codes \textsc{FIREFLY} (e.g., \cite{Comparat2017}) and \textsc{Prospector} (\cite{Leja2017}). For a description of the derivation of stellar properties, we refer the reader to \cite{Ojarzun2019}.
To make a comparison with simulations, we extracted around 2800 MaNGA-like ETGs from the $z=0.1$ snapshot of IllustrisTNG100.
For each simulated galaxy, we choose randomly a line of sight and we build a 2D map of stellar properties by projecting the positions of stellar particles onto a $300\times300$ pixel area. The 1D profiles are derived from the 2D maps using concentric elliptic annuli with fixed ellipticity for each ETG. A detailed description of this fit procedure is provided in \cite{Huang2018}. 

\begin{figure}[b]
    \centering
        \includegraphics[width=1\textwidth]{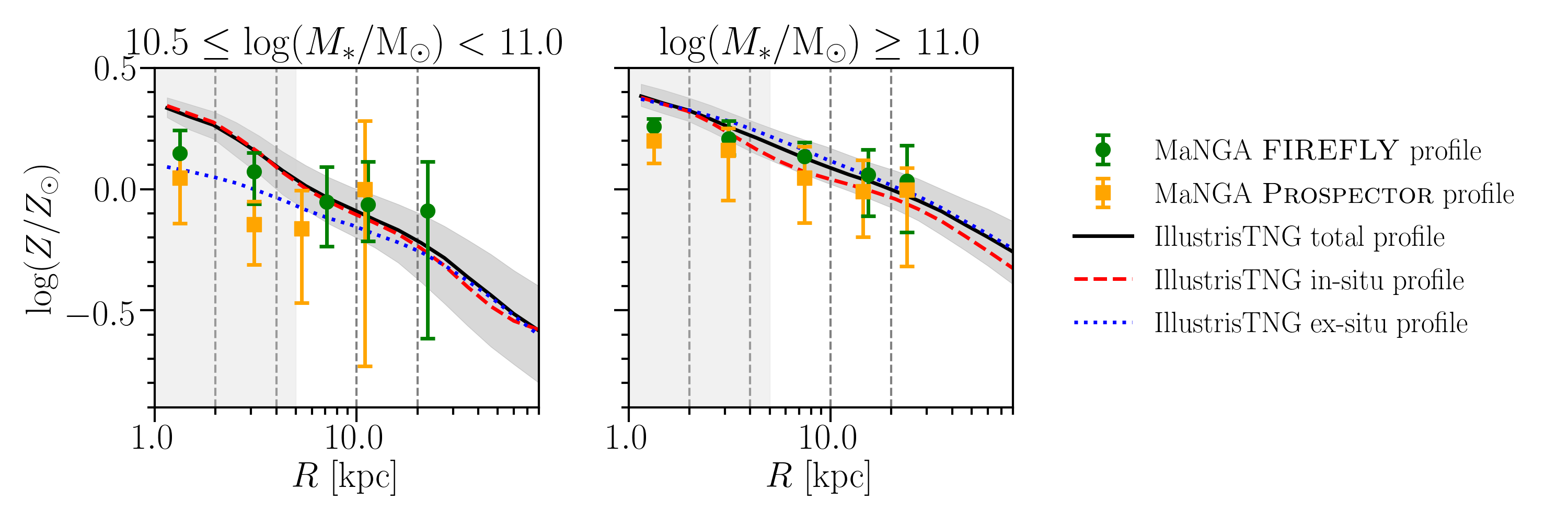}
    \caption{Mass-weighted metallicity radial profiles of massive ETGs with lower (left panel) and higher (right panel) stellar mass. The circles and squares represent the median estimates for MaNGA galaxies from \textsc{FIREFLY} and \textsc{Prospector} codes, respectively.
    The curves represent the median estimates for IllustrisTNG galaxies. The dashed and dotted curves represent the in-situ and the ex-situ stellar components, respectively, while the solid curve represents the total stellar components. The shaded area and the error bars indicate $1\sigma$ scatter.}
    \label{fig:metallicity}
\end{figure}

In \autoref{fig:metallicity}, the median metallicity profiles for MaNGA and IllustrisTNG ETGs are shown for two stellar mass bins. Although the \textsc{Prospector} metallicity tends to be lower than the \textsc{FIREFLY} estimate (offset mainly due to different stellar models assumed), we find good agreement between the two measurements. Moreover, the observed profiles are well reproduced by the IllustrisTNG profiles, in particular at the high-mass end.
In the case of IllustrisTNG galaxies, we can disentangle the in-situ and ex-situ stellar populations (see \cite{RodriguezGomez2016}) and measure for each component its metallicity profile.
Looking at the behaviour of the in-situ and ex-situ metallicity distributions, we notice that for galaxies with $M_*\lesssim10^{11}\,\mathrm{M_\odot}$ the inner regions are  dominated by the in-situ component: the total and the in-situ  metallicity profiles are indistinguishable out to $\mathrm{20\,kpc}$.
Instead, in ETGs with $M_*\gtrsim10^{11}\,\mathrm{M_\odot}$, the ex-situ component is as relevant as (or even more relevant than) the in-situ component at all radii, and has, on average, higher metallicity. 
These results, combined with the finding that the stellar surface density profiles of ETGs with $M_*\gtrsim10^{11}\,\mathrm{M_\odot}$ are similar for in-situ and ex-situ stars (\cite{ChowdhuryInPrep}), are consistent with the fact that major mergers are important in the assembly of the most massive galaxies in  IllustrisTNG (\cite{Tacchella2019}). As already shown in previous works (see \cite{Pillepich2018b}), the role of the ex-situ population tends to be stronger in galaxies with $M_*\gtrsim10^{10.5}\,\mathrm{M_\odot}$, constituting more than the $50\%$ of the total stellar mass.

\section{Implications}
The evolution of the $M_*$-$\sigma_{\rm e}$ relation and the metallicity profiles of massive ETGs can be interpreted in the context of a merger-induced evolution.
The stellar mass of galaxies varies with cosmic time mainly as a consequence of star formation and accretion of stars. On the one hand, some internal mechanisms, such as stellar or active galactic nucleus feedback, can blow out part of the material, depriving the galaxy of the gas reservoir needed to form new stars and then limiting the growth of the stellar mass. On the other hand, phenomena like mergers can trigger star formation and bring in stars formed in other galaxies. In massive systems, like the ETGs considered in these works, the latter process is expected to be dominant.
The results here presented underline the importance of having large and high-resolution observational surveys and cosmological simulations, both necessary to improve our understanding of the galaxy evolution. A self-consistent comparison between observations and simulations is crucial to draw robust conclusions.

\end{document}